\documentclass[prb,aps,twocolumn,showpacs,preprintnumbers,amsmath,amssymb,superscriptaddress,unsortedaddress,floatfix]{revtex4}

\usepackage[dvips]{graphicx}
\usepackage{dcolumn}
\usepackage{bm}

\newcommand{\TN}{\ensuremath{T_{\mathrm{N}}}}

\newcommand{\kkk}{\ensuremath{\langle kkk \rangle}}

\renewcommand{\vec}[1]{\ensuremath{\bm{#1}}}

\begin{document}


\title{Multi-\vec{k} magnetic structures in USb$_{0.9}$Te$_{0.1}$ and UAs$_{0.8}$Se$_{0.2}$
    observed via resonant x-ray scattering at the U M$_4$ edge}

\author{B. Detlefs}
\email{janousova@esrf.fr}
\affiliation{%
European Commission, JRC, Institute for Transuranium Elements, 
Postfach 2340, Karlsruhe, D-76125, Germany}%
\affiliation{European Synchrotron Radiation Facility, BP 220, F-38043 
Grenoble, CEDEX, France}

\author{S. B. Wilkins}%
\altaffiliation[Present address: ]{Brookhaven National Laboratory, Physics Department, Bldg \#510B
Upton, NY, 11973-5000, USA}
\affiliation{%
European Commission, JRC, Institute for Transuranium Elements, 
Postfach 2340, Karlsruhe, D-76125, Germany}%
\affiliation{European Synchrotron Radiation Facility, BP 220, F-38043 
Grenoble, CEDEX, France}

\author{P. Javorsk\'{y}}
 \affiliation{Charles University in Prague, Faculty of Mathematics 
and Physics, Department of Condensed Matter Physics,   Ke Karlovu 5, 121 16 Prague 
2, Czech Republic}%

\author{E. Blackburn}
\altaffiliation[Present address: ]{Dept. of Physics, Univ. of California, San Diego, USA}
 \affiliation{European Commission, JRC, Institute for Transuranium Elements, 
Postfach 2340, Karlsruhe, D-76125, Germany}%

\author{G. H. Lander}
\affiliation{%
European Commission, JRC, Institute for Transuranium Elements, 
Postfach 2340, Karlsruhe, D-76125, Germany}%

\date{\today}

\begin{abstract}
Experiments with resonant photons at the U \textit{M}$_4$ edge have 
been performed on a sample of USb$_{0.9}$Te$_{0.1}$, which has an incommensurate 
magnetic structure with $\vec{k} = k = 0.596(2)$ 
reciprocal lattice units. The reflections of the form $\kkk$, 
as observed previously in a commensurate $k = 1/2$ system [N. 
Bernhoeft \textit{et al}., Phys. Rev. B \textbf{69} 174415 (2004)] are observed, 
removing any doubt that these occur because of multiple scattering 
or high-order contamination of the incident photon beam. They 
are clearly connected with the presence of a $3\vec{k}$ configuration. 
Measurements of the $\kkk$  reflections from the sample 
UAs$_{0.8}$Se$_{0.2}$ in a magnetic field show that the transition at $T^* 
{\sim} 50$~K is between a low-temperature $2\vec{k}$ and 
high-temperature $3\vec{k}$ state and that this transition is 
sensitive to an applied magnetic field. These experiments stress 
the need for quantitative theory to explain the intensities of 
these $\kkk$ reflections.

\end{abstract}

\maketitle

\section{\label{sec:Introduction}Introduction}

In determining magnetic structures for a vast array of materials, 
neutron diffraction is the technique of choice and has been used 
with great success for the last 50 years. However, the realisation 
by Kouvel and Kasper~\cite{Kouvel1963} that multi-$\vec{k}$ configurations, 
in their case in (Ni, Fe)$_{3}$Mn alloys, could be present, often 
makes the ground-state determination by neutron diffraction ambiguous. 
Single-$\vec{k}$ configurations have domains with a single propagation 
direction within a domain, whereas in the multi-$\vec{k}$ configurations 
several propagation directions, commonly three in cubic systems 
but possibly more, exist \emph{simultaneously}.\cite{RossatMignod1987, Forgan1989} The common 
method to examine whether configurations are single- or multi-$\vec{k}$
is to apply an external perturbation (usually magnetic field 
cooling) and to measure the intensities from various ``domains''. 
If they are changed by the perturbation then the configuration 
can often be suggested,\cite{RossatMignod1987, Burlet1986} whereas if they 
are not, a multi-$\vec{k}$ configuration is assumed. Unfortunately, 
one can also envisage that the field cooling itself changes the 
ground state of the system or that the field used is not large 
enough to change the domain population. Thus, the confusion of 
whether a magnetic structure is multi-$\vec{k}$ or not, continues 
to be a limitation of neutron studies of magnetic materials.

With this background it was therefore of considerable interest 
when additional diffraction peaks observed in resonance x-ray 
scattering (RXS), apparently associated uniquely with the $3\vec{k}$
configuration, were reported for a sample of UAs$_{0.8}$Se$_{0.2}$.\cite{Bernhoeft2004}
 These diffraction peaks were weak (less than 10$^{-4}$ of the 
main Bragg peaks arising from magnetic order) but clearly observable 
in the RXS experiments because of the large resonant enhancement 
that occurs when the incident photon energy is tuned to the $M_4$ 
edge of uranium.\cite{McWhan1990} More recently, these extra reflections have 
been observed also using neutron diffraction,\cite{Blackburn2006} confirming 
that the reflections are indeed magnetic dipole in origin. In 
addition, the neutrons demonstrate that the effect is not connected 
with the surface, which is always a possibility if effects are 
observed only with the RXS technique. 

The new reflections are associated with the phase coherence of 
the three different propagation directions that make up the $3\vec{k}$
configuration.\cite{Lander2004} Such reflections were earlier reported in 
the assumed $3\vec{k}$ structure of CeAl$_{2}$.\cite{Shapiro1979} In this latter 
paper, the term giving rise to the $\kkk$  reflections, 
which is a convenient shorthand notation for this subset of reflections 
since they contain a contribution from each $\vec{k}$ component, 
was identified with a $4^{th}$-order term in the free-energy expansion. 
The term is proportional to $M_xM_yM_zM_c$, where 
the individual components of the magnetization propagating along 
the three perpendicular axes are $M_x$ etc, and the term $\vec{k}_c = \vec{k}_x + \vec{k}_y + \vec{k}_z$ 
represents the \emph{coherent} part of the different magnetisation 
distributions. Note that expressing such terms within free energy 
fulfills the symmetry of the system, but gives no idea of their 
intensity with respect to the main $\langle k00\rangle$ magnetic 
Bragg diffraction peaks. However, further research on CeAl$_{2}$ 
established that these reflections came from a different magnetic 
configuration in the system,\cite{Barbara1980} and our present understanding 
of CeAl$_{2}$ is that it is \emph{not} a $3\vec{k}$ system,\cite{Forgan1990, Harris2006} 
so these reflections should not be present. To our knowledge 
this makes the present series of experiments~\cite[and this 
work]{Bernhoeft2004, Blackburn2006} the first to make these observations. Despite doubts of 
the validity of the experimental results in Ref.~\onlinecite{Shapiro1979}, the theoretical 
expression in this reference still gives the correct symmetry 
considerations for the $\kkk$  reflections.

The present paper brings further experimental evidence about 
these unusual diffraction peaks. So far, studies of the $\kkk$
peaks~\cite{Bernhoeft2004, Blackburn2006} have been confined to samples with the commensurate 
magnetic wave vector $k = 1/2$. Although 
significant efforts were made in Refs.~\onlinecite{Bernhoeft2004} and \onlinecite{Blackburn2006} to eliminate multiple 
scattering and higher-order diffraction effects, one way to ensure 
that such effects are further minimised is to examine a material 
in which the magnetic ordering is \emph{incommensurate}. Exploiting 
the magnetic properties found in the system USb--UTe,\cite{Burlet1980} the 
first part of this paper presents clear evidence for the new 
diffraction peaks in such an incommensurate magnetic system.

In the second part of the paper, we return to the commensurate 
system UAs$_{0.8}$Se$_{0.2}$ with $k = 1/2$ and examine the behavior 
of the $\kkk$  reflections as a function of magnetic 
field. Bulk techniques (specific heat and magnetic susceptibility) 
performed on the same sample of UAs$_{0.8}$Se$_{0.2}$ have shown anomalies 
associated with what is believed to be the transition between 
a $2\vec{k}$ state at low temperature and a $3\vec{k}$ state 
at higher temperature~\cite{Bernhoeft2004, Kuznietz1987} and our goal was to study the field-dependence 
of the new diffraction peaks to determine the exact nature of 
the transition. The experiments show that the transition as a 
function of applied field is indeed one between the $2\vec{k}$ 
and $3\vec{k}$ states. 

\section{\label{sec:Experimental}Experimental details and discussion}

All RXS experiments reported here have been performed on the 
ID20 beamline at the European Synchrotron Radiation Facility 
(ESRF), Grenoble, France~\cite{webID20} using Si(111) horizontally focusing 
double crystal monochromator. Two Si-coated mirrors assure vertical 
focusing at the sample position and elimination of higher harmonics 
from the incident beam. Au(111) single crystal was chosen as 
a polarization analyser based on the Bragg diffraction close 
to the Brewster's angle of $\theta = 45{^\circ}$ at the energy of U $M_{4}$ 
edge $E = 3.724$~keV). 

In the case of USb$_{0.9}$Te$_{0.1}$, the sample was installed in a 
closed-cycle cryostat with base temperature of about 10 K, in 
the vertical scattering geometry. This setup, with $[001]$ direction 
vertical, allows for partial rotation of the sample around the 
scattering vector, necessary for azimuthal scans. The UAs$_{0.8}$Se$_{0.2}$ 
experiment was performed in the horizontal scattering geometry, 
with the sample mounted in a 10 T cryomagnet. The magnetic field 
was applied along the $[1\bar{1}0]$ direction allowing reflections 
within the (${h h l}$) plane to be measured. We used polarization 
analysis of the scattered beam only in the first part of the 
experiment to confirm that our signal was uniquely in the rotated 
polarization channel ($\pi \rightarrow \sigma$, where $\pi$ and $\sigma$ 
are the beam polarization components perpendicular and parallel 
to the scattering plane). The main part of the experiment was 
performed without analysis of the scattered beam polarization 
for intensity reasons because the $\kkk$ peak intensity is further reduced due to the poor transmission through 
the Be windows of the cryomagnet. Fortunately, for the scattering 
vector of interest, $\vec{Q} = (1/2\;\; 1/2\;\; 5/2)$, the scattering 
angle $2\theta$ is close to $97^\circ$ which means that with incident beam $\pi$ polarization, 
any background due to a lattice contribution (multiple scattering 
etc.), corresponding to a $\pi\pi$ signal, is negligible.

The specific heat was measured on an 83 mg sample using the relaxation
method in a PPMS-9 equipment (Quantum Design) within the temperature
range 2 -- 300 K. The external magnetic field up to 9~T was applied along the [001] direction.

\subsection{\label{sec:USbTe}Experiments on USb$_{0.9}$Te$_{0.1}$}

The magnetic phase diagram of the USb-UTe solid solutions was 
first examined 25 years ago.\cite{Burlet1980} USb is a well-known (cubic) 
$3\vec{k}$ antiferromagnet with wavevector $\langle k00 \rangle$ 
where $k = 1$ and $\TN = 215$~K. With a small doping of Te 
the magnetic wave vector ($k$) starts to reduce and, at least 
near $\TN$, incommensurate magnetic order is found.\cite{Burlet1980} For our 
experiments we have used a large crystal (because neutron experiments 
were also performed on this sample) of 0.512~g, which has a good 
mosaic of 0.026 degree full-width at half maximum. 

\begin{figure}
\includegraphics[width=0.8\columnwidth]{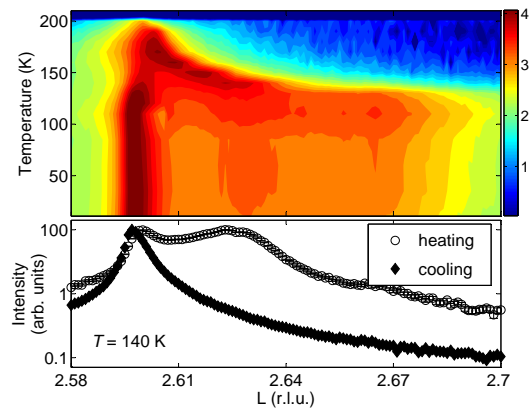}
\caption{\label{fig1} (Color online) Upper panel: Magnetic intensity from the sample of USb$_{0.9}$Te$_{0.1}$ as a function of scattering vector $(00L)$ and temperature.  The sample was first cooled to 10~K and then data collected as the sample was heated. Lower panel: Data at $T = 140$~K on both heating and on cooling. The main magnetic modulation corresponds to $k \sim 0.6$~r.l.u., originating from the $(002)$ charge reflection, but intensity is observed over a wide range of wave vector. At the highest temperatures, just below $\TN \sim 200$~K, a single incommensurate magnetic wave vector is found, in agreement with the phase diagram proposed in Ref.~\onlinecite{Burlet1980}. Scale in both panels is logarithmic.}
\end{figure}

\begin{figure}
\includegraphics[width=0.7\columnwidth]{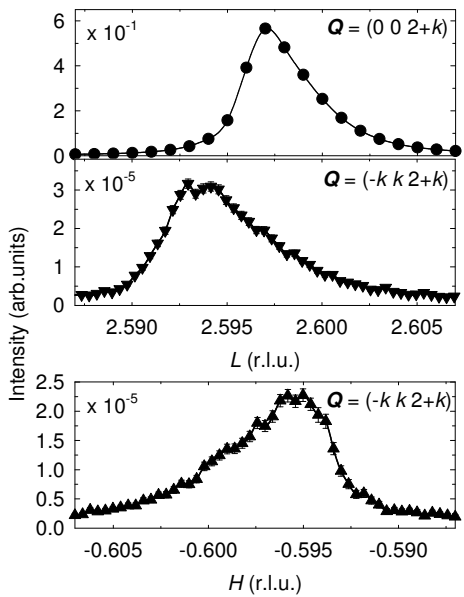}
\caption{\label{fig2}Comparison of intensity profiles of a $\langle k00\rangle$ peak (upper panel, as in Fig.~\ref{fig1}) and $\kkk$ peak (lower two panels) at $T = 140$~K. Notice that the asymmetry extends to higher $k$-value on both $\langle k00 \rangle$  and $\kkk$ peaks and in both $L$-scan and $H$-scan. A small difference in the actual value of $k$ for $\langle k00 \rangle$ and $\kkk$ peaks is due to crystal/diffractometer misalignment. The true value of $k$ is determined from the positions of the third harmonics reflections $(0\; 0\; 3k)$ and $(0\; 0\; 4-3k)$ to be 0.596(2) r.l.u. The $\langle k00 \rangle$ peak has a full-width at half maximum (FWHM) of 0.0036 r.l.u. in the $L$ direction and 0.0032 r.l.u. in the transverse ($H$) direction.  This may be compared to 0.0023 and 0.0014 r.l.u., respectively, for the charge peaks measured at the same photon energy. The latter are close to the instrumental resolution, which is poor in the chi direction ($K$). The $\kkk$ peaks have FWHM's of about 0.0060 r.l.u in both the $L$ and $H$ directions.}
\end{figure}

An overview of the magnetic scattering as a function of temperature 
is given in Fig.~\ref{fig1}. The scattering arises from magnetic modulations 
with wave vectors ($k$) extending from $\sim\:0.6$ to $\sim\: 0.7$ 
reciprocal lattice units (r.l.u.). The pattern shows a more complicated 
situation than reported earlier.\cite{Burlet1980} Presumably, this may be 
explained by the poorer resolution of the early neutron experiments 
as well as the much larger intensity, and thus sensitivity, obtained 
in the RXS technique. All intensity in this plot is measured 
in the $\sigma \rightarrow \pi$ channel, and since there 
is no scattering in the $\sigma \rightarrow \sigma$ channel, 
the scattering is magnetic in origin. At high temperatures, near \TN, 
the scattering is incommensurate and consists of one modulation 
with ${k} {\sim} 0.6$~r.l.u, as reported,\cite{Burlet1980} but below about 
160 K the material exhibits frustration with magnetic scattering 
distributed over a wide range of modulation vectors. In addition 
to this extension of the modulation along the longitudinal direction 
$[0 0 L]$ (as shown in Fig.~\ref{fig1}), the scattering in the transverse 
direction is broad in $q$ as well, showing the short-range 
ordering of the magnetic correlations. Figure~\ref{fig1} shows that at 
all temperatures there is a strong and relatively sharp component 
at $k = 0.6$, and this, rather than a modulation at $k = 
2/3 = 0.667$~r.l.u., is clearly the dominant correlation in the 
system. Moreover, by slow cooling of the sample it was found 
that the $k \sim 0.6$~r.l.u. modulation was maintained 
to at least 140 K. An accurate measure of ${k} = 0.596(2)$~r.l.u. 
was made at this temperature. All subsequent experiments were 
performed at 140 K.

On cooling into the ordered state, no sign of any \emph{external} 
structural distortion (i.e. a distortion from cubic to tetragonal) 
was observed. This is consistent with the fact that all the antiferromagnet 
configurations in the USb--UTe solid solutions are $3\vec{k}$
in nature.\cite{Burlet1980}

In agreement with previous work on the UAs--USe solid solutions, 
we found the peaks referred to as $\langle {kk}0\rangle$.\cite{Longfield2002} These 
arise from the E1-F$^{[2]}$ term in the RXS cross section~\cite{Hill1996} and 
are commonly associated with the quadrupole interactions in these 
materials. They are characterised by a sharp photon energy dependence 
of the scattering, intensity in both $\sigma {\rightarrow} {\pi}$ 
and ${\sigma} {\rightarrow} {\sigma}$ channels, and a characteristic 
azimuthal dependence of the scattered intensity.\cite{Longfield2002}

The reflections characterising the $3\vec{k}$ magnetic structure 
have the form $\kkk$~\cite{Bernhoeft2004} and we show two such reflections 
in Fig.~\ref{fig2}. Comparison of the intensities between data in 
Fig.~\ref{fig2} gives approximately a factor of (3x10$^{-5}$/0.6) x (0.0060/0.0032)$^{2}$ 
= 1.7 x 10$^{-4}$, where the first term is simply the intensity 
of the peak and the second factor takes into account the different 
widths in the $H$ and $L$ directions (the resolution in the $K$ direction 
of the spectrometer is poor and integrates all the signal). This 
is consistent with previous estimates.~\cite{Bernhoeft2004} It is noteworthy that 
the correlations between the different components in the $3\vec{k}$ 
structure give rise to a correlation length that is almost double 
that of a single $\vec{k}$ component. This was not previously 
observed~\cite{Bernhoeft2004} because for a \emph{commensurate} system (like in the 
UAs--USe) the strain terms locking the magnetic and charge modulations 
in the free energy ensure that the magnetic correlations are 
as long range as the charge (see Fig. 2 of Ref.~\onlinecite{Bernhoeft2004}). However, 
this is not the case in an incommensurate system as examined 
here. 

As reported earlier~\cite{Bernhoeft2004}, all the signal for the $\kkk$
peaks was in the ${\sigma} \rightarrow{\pi}$ channel and 
the photon energy dependence was similar to that of the $\langle {k}00\rangle$
peak, i.e. a Lorentzian centered at the U $M_4$ resonance. 
This assures there is no charge contribution to these reflections. 

\begin{figure}
\includegraphics[width=\columnwidth,clip=true]{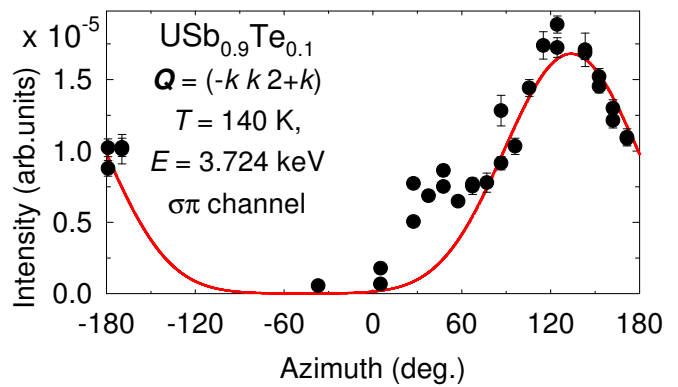}
\caption {\label{fig3} (Color online) Azimuthal scan of the $(-k k 2+k)$ reflection. The solid line is a fit to model in which the dipole moment for the reflection rotates around the direction $[-1 1 1]$.}
\end{figure}

An azimuthal scan of the same reflection is shown in Fig.~\ref{fig3}.
 The dipole component from which this reflection arises is 
$[-1 1 1]$ as shown by the simulation of the azimuthal dependence 
from such a component, scaled to the maximum observed value and 
with no other variable. It should be realized that the azimuthal 
dependence of the $\langle{k}00\rangle$ reflections corresponds 
to a dipole moment along a $\langle 100\rangle$ direction; thus the 
$\kkk$ reflections, which have their dipole moment 
corresponding to one of the $\langle 111 \rangle$ directions, show the 
essential $3\vec{k}$ nature of the configuration. This is discussed 
in Ref.~\onlinecite{Blackburn2006}, where a similar azimuthal dependence was found in 
the USb$_{0.85}$Te$_{0.15}$ material.

\begin{figure}
\includegraphics*[width=\columnwidth]{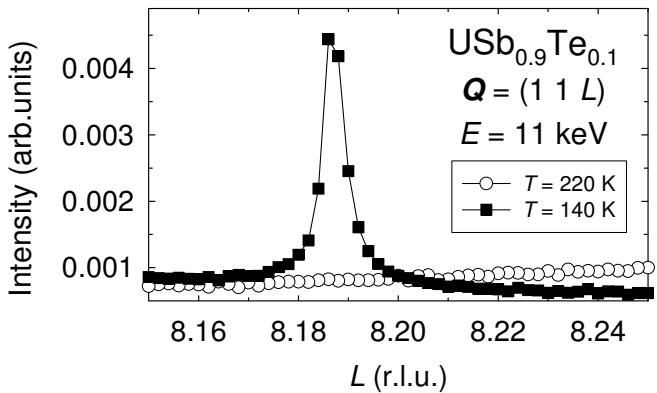}
\caption {\label{fig4} Scan along $[1 1 L]$ direction with the photon energy at 11~keV and a Ge(111) analyser installed in the PA stage. The very sharp peak at $(1\; 1\; 7+2k)$ corresponds to the magneto-elastic term. It gives an excellent measure of the magnitude of $k$.}
\end{figure}

Finally, as pointed out first by McWhan \textit{et al.},\cite{McWhan1990} and 
later in more detail in the UAs--USe system by Longfield \textit{et 
al.},\cite{Longfield2001a} there should be a magneto-elastic distortion in these 
systems corresponding to peaks at $\langle2{k} 0 0\rangle$. This 
is a charge term corresponding to the charge-density wave riding 
on the underlying magnetic modulation. The intensity of this 
peak increases as $\vec{Q}^{2}$, where $\vec{Q}$ is the scattering 
vector. Since the displacement-wave amplitude is small, both 
the penetration and limited $Q$ range available at 3.7~keV mitigate 
against the observation of such peaks. Thus, the photon energy 
was increased to 11~keV and a Ge(111) crystal mounted in the 
polarisation stage. A resultant scan over along the direction 
$[1 1 L]$ is shown in Fig.~\ref{fig4}. This clearly shows the magneto-elastic 
$2k$ component and, as expected, that it exists in an incommensurate 
system.

In summary, for the USb$_{0.9}$Te$_{0.1}$ we emphasize that the magnetic 
modulation is \emph{incommensurate}, but that the $\kkk$ 
peak exists at the same level of intensity ($ \sim 10^{-4}$) 
as compared to the standard $\langle {k}00\rangle$ magnetic reflections. 
These peaks as seen by x-rays arise from the phase coherence 
between the different components in the $3\vec{k}$ structure 
and are not artefacts of the data collection or reflections arising 
from higher-order components in the photon beam.

\subsection{\label{sec:UAsSe}Experiments on USb$_{0.9}$Te$_{0.1}$}

\begin{figure}
\includegraphics*[width=\columnwidth]{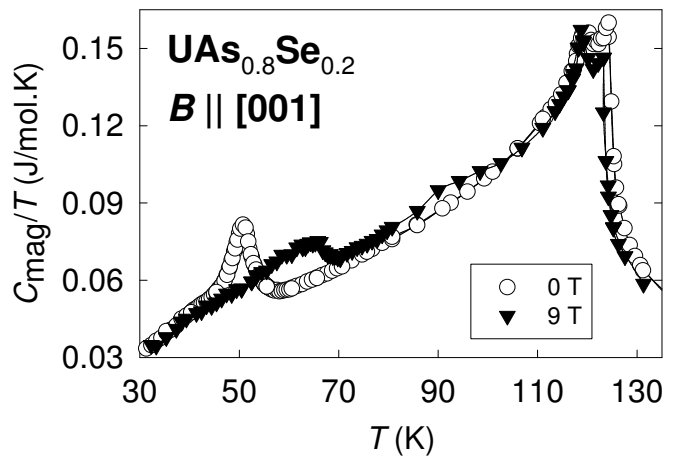}
\caption {\label{fig5} Specific heat measured on sample I of UAs$_{0.8}$Se$_{0.2}$ as a function of magnetic field. The ordering at $\TN \sim 125$~K is insensitive to magnetic field, but the transition ($T^*$) at $\sim  50$~K increases with increasing $B$.}
\end{figure}

In the previous work on this system,\cite{Bernhoeft2004, Blackburn2006} it was shown that 
the $\kkk$ reflections appeared to disappear (on 
cooling) below $T^*$ (${\sim} 50$~K) and thus corresponds to 
a transition between the high-temperature $3\vec{k}$ to the lower 
temperature $2\vec{k}$ state, in which state the reflections 
should not exist, as originally proposed by Kuznietz \textit{et al}. 
\cite{Kuznietz1987} Bulk measurements of specific heat and magnetisation showed 
anomalies also at this temperature.\cite{Bernhoeft2004} An overview of the specific-heat 
data with the original sample (I) is shown in Fig.~\ref{fig5}.

\begin{figure}
\includegraphics*[width=0.7\columnwidth]{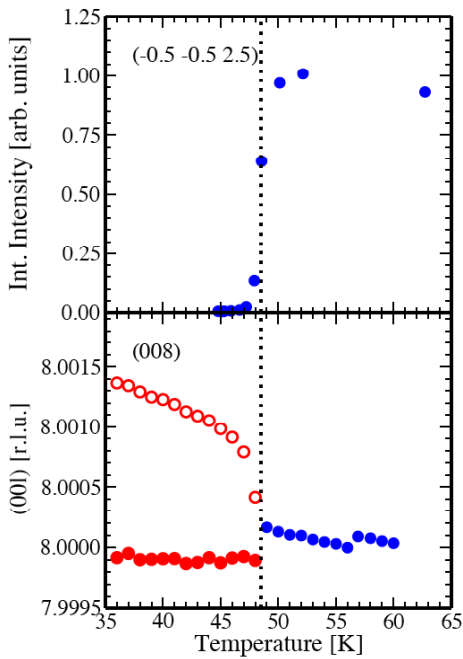}
\caption {\label{fig6}(Color online)  Upper panel: Intensity of the $(-0.5\; 0.5\; 2.5)$ reflection from sample II on heating from low temperature and with a photon energy tuned to the U $M_4$ resonance.
Lower panel: Position of the lattice peak as measured with high resolution $8$~keV photons + Ge(111) analyser. A double peak is observed below $T^*$ indicating an external lattice distortion to tetragonal (or lower) symmetry.
}
\end{figure}

With synchrotron experiments on a new sample of UAs$_{0.8}$Se$_{0.2}$ 
(sample II) we have first re-examined the nature of the transition 
at ${T}^*$. Figure~\ref{fig6} shows (upper panel) the temperature dependence 
of a $\kkk$ reflection and (lower panel) the behavior of 
the lattice as measured with high-resolution photons. Below ${T}^*$, 
there is an external lattice distortion to at least a tetragonal 
symmetry (and may be lower) and this results in a splitting of 
the (008) charge reflection. These results are consistent with 
the idea that the material transforms on cooling from the high-temperature 
$3\vec{k}$ state, which has cubic symmetry, to the low-temperature 
$2\vec{k}$ state, which has lower than cubic symmetry.

To compare the data taken with specific heat and RXS, we have 
chosen to normalize the data to $T^*$ as the specific heat were 
taken on sample I and RXS measurements on sample II. Small differences 
in Se composition may account for a small difference in ${T}^*$ 
of about 3 K; compare Figs.~\ref{fig5} and \ref{fig6}. Furthermore, the original 
specific-heat data were taken with $B \parallel [001]$, 
whereas the RXS measurements with $B \parallel [1 \bar{1} 0]$. 
The latter geometry is required to obtain the $\kkk$ 
peak in the equatorial plane of the magnet. However, the point 
of this study is to show that the change in the intensity of 
the $\kkk$ reflection as a function of applied field 
follows (approximately) the phase transition, and thus identifying 
it as one between $2\vec{k}$ ($T  < {T}^*$) and $3\vec{k}$ 
(${T} >{T}^*$), rather than to understand the full ($B,T$) 
phase diagram of this material. 

\begin{figure}
\includegraphics*[width=\columnwidth]{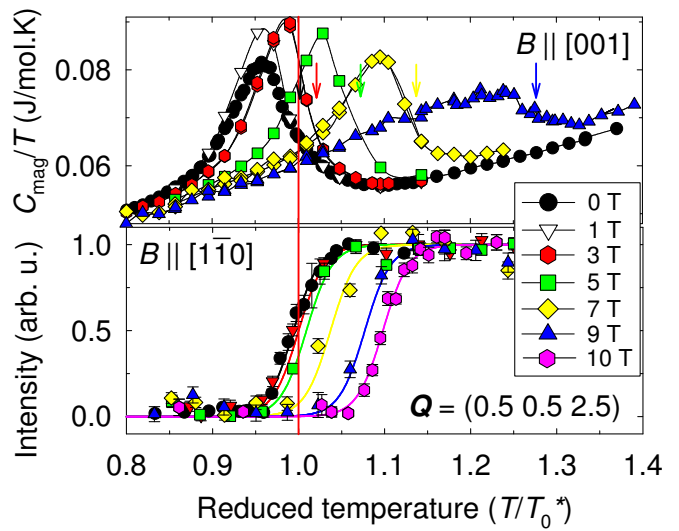}
\caption {\label{fig7}(Color online)  Specific heat (upper part) on sample I divided by $T$ and with the phonon contribution subtracted and (lower part) intensity of the $(0.5\; 0.5\; 2.5)$ reflection in sample II of UAs$_{0.8}$Se$_{0.2}$ as a function of temperature normalized to $T^*$ at $B=0$. The arrows in the upper panel indicate $T^*$. Notice that the magnetic fields have been applied in two different directions for the different techniques.}
\end{figure}

Figure~\ref{fig7} (upper part) shows the change of specific heat (divided 
by ${T}$ and with the phonon contribution subtracted) and the 
intensity of a $\kkk$ reflection as a function of ${T}/{T}^*_{B=0}$ 
for different applied magnetic fields. In addition to the fact 
that both measurements show $T^*$ rising as a function of applied 
field, there are two further points of note in this figure.

The first is that the rise of ${T}^*$ is much greater for $\vec{B} \parallel [001]$ 
than $\vec{B} \parallel [1 \bar{1} 0]$. Since the phase diagram as 
a function of applied field has not been reported for UAs$_{0.8}$Se$_{0.2}$, 
our knowledge is necessarily limited. However, studies of UAs~\cite{RossatMignod1987, Burlet1986, Felcher1979} and UAs$_{0.75}$Se$_{0.25}$~\cite{Kuznietz1990} give some indication what 
might occur. In both cases the easy axis is not $\langle 100 \rangle$, 
but the initial transitions occur at lower fields (at constant 
temperature) with $\vec{B} \parallel [001]$ than any other direction. 
So the transition at ${T}^*$ is more rapidly shifted with $\vec{B} \parallel [001]$ 
than $\vec{B} \parallel [1 \bar{1} 0]$, as observed in Fig.~\ref{fig7}. Further 
details of these effects as a function of magnetic field direction 
may be found in Ref.~\onlinecite{Vogt1993}.

The second observation is related to the surprising result for 
the specific heat for $B = 9$~T, which shows a broad peak extending 
over many degrees. In contrast, the breadth of the transition, 
over which the $\kkk$ reflection develops for high 
fields, is about the same as at lower fields. Although, as in 
the first part, we have not done sufficient experiments to verify 
the exact nature of the high-field state, it seems likely that 
a second transition occurs for fields near 9~T and above. Again 
taking the work on UAs\cite{RossatMignod1987, Burlet1986, Felcher1979, Vogt1993} and UAs$_{0.75}$Se$_{0.25}$~\cite{Kuznietz1990} 
as a guide, a transition to a modified magnetic structure occurs 
for these high fields. With $\vec{B} \parallel [001]$  the transition 
involves a change in the modulation component parallel to $\vec{B}$,\cite{Felcher1979} but this may not be relevant with $\vec{B} \parallel [1 \bar{1} 0]$. Perhaps the surprising result is not that the specific 
heat anomaly at these high fields is so smeared over many degrees, 
but that the intensity of the $\kkk$ reflection shows 
clearly that whatever the magnetic state is above $T^*$, it 
is certainly $3\vec{k}$ in nature.


These results shown in Fig.~\ref{fig7} leave little doubt that the transition 
at $T^*$ is between the $2\vec{k}$ (for $T<{T}^*$) and 
$3\vec{k}$ (for $T >{T}^*$) states, with the $2\vec{k}$ 
state being stabilized with increasing magnetic field. 

\section{\label{sec:Conclusions}Conclusions}

Although the earlier papers in this series~\cite{Bernhoeft2004, Blackburn2006} provided strong 
evidence that the new $\kkk$ peaks were associated 
with the $3\vec{k}$ nature of the magnetic configuration, some 
doubt remained. In particular, the observation of very weak reflections 
at highly symmetric lattice points must always be regarded with 
caution as complicated effects from higher-order components in 
the incident photon or neutron beam or multiple scattering may 
always produce weak reflections at these special positions. We 
are dealing with intensities of the order of 10$^{-4}$ of the magnetic 
peaks, which, in the case of neutrons or RXS at the U $M$ edges, 
represent an intensity of ${\sim} 10^{-6}$ of the nuclear or 
charge peaks, respectively. The present experiments on a $3\vec{k}$ \emph{incommensurate} 
system (USb$_{0.9}$Te$_{0.1}$) show unambiguously that the $\kkk$ 
reflections exist and behave in the same way as previously reported.\cite{Bernhoeft2004}
 They clearly arise from the coherence of the individual 
three components in the system.

Furthermore, a careful examination of the transition at $T^*$ 
in UAs$_{0.8}$Se$_{0.2}$, which had been proposed\cite{Kuznietz1987} as a transition 
between the $2\vec{k}$ (low temperature) and $3\vec{k}$ (high 
temperature) states, shows that the field dependence of the $\kkk$ 
reflection confirms this proposal. As the field is increased, 
the $2\vec{k}$ state is stabilized at the expense of the $3\vec{k}$
modulations.

The observation and explanation of these $\kkk$ reflections 
in terms of the free energy, as proposed\cite{Shapiro1979} in connection with 
CeAl$_{2}$, is satisfying from a conceptual point of view as to 
what symmetry considerations define these $\kkk$
peaks. An alternative analysis in terms of Clifford algebra for 
the presence of these peaks has also been recently advanced.\cite{Blackburn2006a} 
In neither case do the arguments specify the intensity to be 
expected. New experiments that we are undertaking are to measure 
the form factor, $f(Q)$, of the $\kkk$ reflections 
with neutron scattering so as to determine the spatial distribution 
of the magnetization that is involved in the coherence between 
the different $3\vec{k}$ components.

More than 40 years after the discovery of multi-$\vec{k}$ magnetic 
structures,\cite{Kouvel1963} some aspects of their nature remain elusive; 
it is surprising, for example, that $\kkk$ reflection 
have not been observed previously. Ironically, multi-$\vec{k}$ 
configurations have been predicted\cite{Lindgard2003} to be the stable magnetic 
states in nano-antiferromagnets, simply because of the high cost 
in energy of domain walls. Perhaps this gives a further motivation 
to fully understanding the subtleties of multi-$\vec{k}$ structures.

\begin{acknowledgments}
We thank Carsten Detlefs, Nick Bernhoeft, and Luigi Paolasini 
for discussions and help with some aspects of the experiments. 
B.D., S.B.W., P.J., and E.B. thank the European Commission for 
support in the frame of the ``Training and Mobility of Researchers'' 
program. Financial support for access to the Actinide User Laboratory 
at ITU-Karlsruhe, in which the specific-heat measurements were 
performed, within the frame of the European Community-Access 
to Research Infrastructures action of the Improving Human Potential 
Program (IHP), contract HPRI-CT-2002-00118, is acknowledged.
\end{acknowledgments}

\bibliography{d:/blanka/diff_papers/ref_database}

\begin{thebibliography}{23}
\expandafter\ifx\csname natexlab\endcsname\relax\def\natexlab#1{#1}\fi
\expandafter\ifx\csname bibnamefont\endcsname\relax
  \def\bibnamefont#1{#1}\fi
\expandafter\ifx\csname bibfnamefont\endcsname\relax
  \def\bibfnamefont#1{#1}\fi
\expandafter\ifx\csname citenamefont\endcsname\relax
  \def\citenamefont#1{#1}\fi
\expandafter\ifx\csname url\endcsname\relax
  \def\url#1{\texttt{#1}}\fi
\expandafter\ifx\csname urlprefix\endcsname\relax\def\urlprefix{URL }\fi
\providecommand{\bibinfo}[2]{#2}
\providecommand{\eprint}[2][]{\url{#2}}

\bibitem[{\citenamefont{Kouvel and Kasper}(1963)}]{Kouvel1963}
\bibinfo{author}{\bibfnamefont{J.~S.} \bibnamefont{Kouvel}} \bibnamefont{and}
  \bibinfo{author}{\bibfnamefont{J.~S.} \bibnamefont{Kasper}},
  \bibinfo{journal}{J. Phys. Chem. Solids} \textbf{\bibinfo{volume}{24}},
  \bibinfo{pages}{529} (\bibinfo{year}{1963}).

\bibitem[{\citenamefont{Rossat-Mignod}(1987)}]{RossatMignod1987}
\bibinfo{author}{\bibfnamefont{J.}~\bibnamefont{Rossat-Mignod}}, in
  \emph{\bibinfo{booktitle}{Methods in Experimental Physics}}, edited by
  \bibinfo{editor}{\bibfnamefont{K.}~\bibnamefont{Skjold}} \bibnamefont{and}
  \bibinfo{editor}{\bibfnamefont{D.~L.} \bibnamefont{Price}}
  (\bibinfo{publisher}{Academic}, \bibinfo{year}{1987}),
  p.~\bibinfo{pages}{69}.

\bibitem[{\citenamefont{Forgan et~al.}(1989)\citenamefont{Forgan, Gibbons,
  McEwen, and Fort}}]{Forgan1989}
\bibinfo{author}{\bibfnamefont{E.~M.} \bibnamefont{Forgan}},
  \bibinfo{author}{\bibfnamefont{E.~P.} \bibnamefont{Gibbons}},
  \bibinfo{author}{\bibfnamefont{K.~A.} \bibnamefont{McEwen}},
  \bibnamefont{and} \bibinfo{author}{\bibfnamefont{D.}~\bibnamefont{Fort}},
  \bibinfo{journal}{Phys. Rev. Lett.} \textbf{\bibinfo{volume}{62}},
  \bibinfo{pages}{470} (\bibinfo{year}{1989}).

\bibitem[{\citenamefont{Burlet et~al.}(1986)\citenamefont{Burlet,
  Rossat-Mignod, Quezel, Vogt, Spirlet, and Rebizant}}]{Burlet1986}
\bibinfo{author}{\bibfnamefont{P.}~\bibnamefont{Burlet}},
  \bibinfo{author}{\bibfnamefont{J.}~\bibnamefont{Rossat-Mignod}},
  \bibinfo{author}{\bibfnamefont{S.}~\bibnamefont{Quezel}},
  \bibinfo{author}{\bibfnamefont{O.}~\bibnamefont{Vogt}},
  \bibinfo{author}{\bibfnamefont{J.~C.} \bibnamefont{Spirlet}},
  \bibnamefont{and} \bibinfo{author}{\bibfnamefont{J.}~\bibnamefont{Rebizant}},
  \bibinfo{journal}{J. Less-Common Met.} \textbf{\bibinfo{volume}{121}},
  \bibinfo{pages}{121} (\bibinfo{year}{1986}).

\bibitem[{\citenamefont{Bernhoeft et~al.}(2004)\citenamefont{Bernhoeft,
  Paix\~ao, Detlefs, Wilkins, Javorsk\'y, Blackburn, and
  Lander}}]{Bernhoeft2004}
\bibinfo{author}{\bibfnamefont{N.}~\bibnamefont{Bernhoeft}},
  \bibinfo{author}{\bibfnamefont{J.~A.} \bibnamefont{Paix\~ao}},
  \bibinfo{author}{\bibfnamefont{C.}~\bibnamefont{Detlefs}},
  \bibinfo{author}{\bibfnamefont{S.~B.} \bibnamefont{Wilkins}},
  \bibinfo{author}{\bibfnamefont{P.}~\bibnamefont{Javorsk\'y}},
  \bibinfo{author}{\bibfnamefont{E.}~\bibnamefont{Blackburn}},
  \bibnamefont{and} \bibinfo{author}{\bibfnamefont{G.~H.}
  \bibnamefont{Lander}}, \bibinfo{journal}{Phys. Rev. B}
  \textbf{\bibinfo{volume}{69}}, \bibinfo{pages}{174415}
  (\bibinfo{year}{2004}).

\bibitem[{\citenamefont{McWhan et~al.}(1990)\citenamefont{McWhan, Vettier,
  Isaacs, Ice, Siddons, Hastings, Peters, and Vogt}}]{McWhan1990}
\bibinfo{author}{\bibfnamefont{D.~B.} \bibnamefont{McWhan}},
  \bibinfo{author}{\bibfnamefont{C.}~\bibnamefont{Vettier}},
  \bibinfo{author}{\bibfnamefont{E.~D.} \bibnamefont{Isaacs}},
  \bibinfo{author}{\bibfnamefont{G.~E.} \bibnamefont{Ice}},
  \bibinfo{author}{\bibfnamefont{D.~P.} \bibnamefont{Siddons}},
  \bibinfo{author}{\bibfnamefont{J.~B.} \bibnamefont{Hastings}},
  \bibinfo{author}{\bibfnamefont{C.}~\bibnamefont{Peters}}, \bibnamefont{and}
  \bibinfo{author}{\bibfnamefont{O.}~\bibnamefont{Vogt}},
  \bibinfo{journal}{Phys. Rev. B} \textbf{\bibinfo{volume}{42}},
  \bibinfo{pages}{6007} (\bibinfo{year}{1990}).

\bibitem[{\citenamefont{Blackburn et~al.}(2006)\citenamefont{Blackburn,
  Bernhoeft, McIntyre, Wilkins, Boulet, Ollivier, Podlesnyak, Juranyi,
  Javorsk\'y, Lander et~al.}}]{Blackburn2006}
\bibinfo{author}{\bibfnamefont{E.}~\bibnamefont{Blackburn}},
  \bibinfo{author}{\bibfnamefont{N.}~\bibnamefont{Bernhoeft}},
  \bibinfo{author}{\bibfnamefont{G.~J.} \bibnamefont{McIntyre}},
  \bibinfo{author}{\bibfnamefont{S.~B.} \bibnamefont{Wilkins}},
  \bibinfo{author}{\bibfnamefont{P.}~\bibnamefont{Boulet}},
  \bibinfo{author}{\bibfnamefont{J.}~\bibnamefont{Ollivier}},
  \bibinfo{author}{\bibfnamefont{A.}~\bibnamefont{Podlesnyak}},
  \bibinfo{author}{\bibfnamefont{F.}~\bibnamefont{Juranyi}},
  \bibinfo{author}{\bibfnamefont{P.}~\bibnamefont{Javorsk\'y}},
  \bibinfo{author}{\bibfnamefont{G.~H.} \bibnamefont{Lander}},
  \bibnamefont{et~al.}, \bibinfo{journal}{Phil. Mag.}
  \textbf{\bibinfo{volume}{86}}, \bibinfo{pages}{2553} (\bibinfo{year}{2006}).

\bibitem[{\citenamefont{Lander and Bernhoeft}(2004)}]{Lander2004}
\bibinfo{author}{\bibfnamefont{G.~H.} \bibnamefont{Lander}} \bibnamefont{and}
  \bibinfo{author}{\bibfnamefont{N.}~\bibnamefont{Bernhoeft}},
  \bibinfo{journal}{Physica B} \textbf{\bibinfo{volume}{345}},
  \bibinfo{pages}{34} (\bibinfo{year}{2004}).

\bibitem[{\citenamefont{Shapiro et~al.}(1979)\citenamefont{Shapiro, Gurewitz,
  Parks, and Kupferberg}}]{Shapiro1979}
\bibinfo{author}{\bibfnamefont{S.~M.} \bibnamefont{Shapiro}},
  \bibinfo{author}{\bibfnamefont{E.}~\bibnamefont{Gurewitz}},
  \bibinfo{author}{\bibfnamefont{R.~D.} \bibnamefont{Parks}}, \bibnamefont{and}
  \bibinfo{author}{\bibfnamefont{L.~C.} \bibnamefont{Kupferberg}},
  \bibinfo{journal}{Phys. Rev. Lett.} \textbf{\bibinfo{volume}{43}},
  \bibinfo{pages}{1748} (\bibinfo{year}{1979}).

\bibitem[{\citenamefont{Barbara et~al.}(1980)\citenamefont{Barbara, Rossignol,
  Boucherle, and Vettier}}]{Barbara1980}
\bibinfo{author}{\bibfnamefont{B.}~\bibnamefont{Barbara}},
  \bibinfo{author}{\bibfnamefont{M.~F.} \bibnamefont{Rossignol}},
  \bibinfo{author}{\bibfnamefont{J.~X.} \bibnamefont{Boucherle}},
  \bibnamefont{and} \bibinfo{author}{\bibfnamefont{C.}~\bibnamefont{Vettier}},
  \bibinfo{journal}{Phys. Rev. Lett.} \textbf{\bibinfo{volume}{45}},
  \bibinfo{pages}{938} (\bibinfo{year}{1980}).

\bibitem[{\citenamefont{Forgan et~al.}(1990)\citenamefont{Forgan, Rainford,
  Lee, Abell, and Bi}}]{Forgan1990}
\bibinfo{author}{\bibfnamefont{E.~M.} \bibnamefont{Forgan}},
  \bibinfo{author}{\bibfnamefont{B.~D.} \bibnamefont{Rainford}},
  \bibinfo{author}{\bibfnamefont{S.~L.} \bibnamefont{Lee}},
  \bibinfo{author}{\bibfnamefont{J.~S.} \bibnamefont{Abell}}, \bibnamefont{and}
  \bibinfo{author}{\bibfnamefont{Y.}~\bibnamefont{Bi}}, \bibinfo{journal}{J.
  Phys. Cond. Matt.} \textbf{\bibinfo{volume}{2}}, \bibinfo{pages}{10211}
  (\bibinfo{year}{1990}).

\bibitem[{\citenamefont{Harris and Schweizer}(2006)}]{Harris2006}
\bibinfo{author}{\bibfnamefont{A.~B.} \bibnamefont{Harris}} \bibnamefont{and}
  \bibinfo{author}{\bibfnamefont{J.}~\bibnamefont{Schweizer}},
  \bibinfo{journal}{Phys. Rev. B} \textbf{\bibinfo{volume}{74}},
  \bibinfo{pages}{134411} (\bibinfo{year}{2006}).

\bibitem[{\citenamefont{Burlet et~al.}(1980)\citenamefont{Burlet, Quezel,
  Rossat-Mignod, Vogt, and Lander}}]{Burlet1980}
\bibinfo{author}{\bibfnamefont{P.}~\bibnamefont{Burlet}},
  \bibinfo{author}{\bibfnamefont{S.}~\bibnamefont{Quezel}},
  \bibinfo{author}{\bibfnamefont{J.}~\bibnamefont{Rossat-Mignod}},
  \bibinfo{author}{\bibfnamefont{O.}~\bibnamefont{Vogt}}, \bibnamefont{and}
  \bibinfo{author}{\bibfnamefont{G.~H.} \bibnamefont{Lander}},
  \bibinfo{journal}{Physica B \& C} \textbf{\bibinfo{volume}{102}},
  \bibinfo{pages}{271} (\bibinfo{year}{1980}).

\bibitem[{\citenamefont{Kuznietz et~al.}(1987)\citenamefont{Kuznietz, Burlet,
  Rossat-Mignod, and Vogt}}]{Kuznietz1987}
\bibinfo{author}{\bibfnamefont{M.}~\bibnamefont{Kuznietz}},
  \bibinfo{author}{\bibfnamefont{P.}~\bibnamefont{Burlet}},
  \bibinfo{author}{\bibfnamefont{J.}~\bibnamefont{Rossat-Mignod}},
  \bibnamefont{and} \bibinfo{author}{\bibfnamefont{O.}~\bibnamefont{Vogt}},
  \bibinfo{journal}{J. Magn. Magn. Mater.} \textbf{\bibinfo{volume}{69}},
  \bibinfo{pages}{12} (\bibinfo{year}{1987}).

\bibitem[{web()}]{webID20}
\emph{\bibinfo{title}{web site of id20}},
  \urlprefix\url{http://www.esrf.fr/exp_facilities/ID20/html/id20.html}.

\bibitem[{\citenamefont{Longfield et~al.}(2002)\citenamefont{Longfield,
  Paix\~ao, Bernhoeft, and Lander}}]{Longfield2002}
\bibinfo{author}{\bibfnamefont{M.~J.} \bibnamefont{Longfield}},
  \bibinfo{author}{\bibfnamefont{J.~A.} \bibnamefont{Paix\~ao}},
  \bibinfo{author}{\bibfnamefont{N.}~\bibnamefont{Bernhoeft}},
  \bibnamefont{and} \bibinfo{author}{\bibfnamefont{G.~H.}
  \bibnamefont{Lander}}, \bibinfo{journal}{Phys. Rev. B}
  \textbf{\bibinfo{volume}{66}}, \bibinfo{pages}{054417}
  (\bibinfo{year}{2002}).

\bibitem[{\citenamefont{Hill and McMorrow}(1996)}]{Hill1996}
\bibinfo{author}{\bibfnamefont{J.~P.} \bibnamefont{Hill}} \bibnamefont{and}
  \bibinfo{author}{\bibfnamefont{D.~F.} \bibnamefont{McMorrow}},
  \bibinfo{journal}{Acta Cryst. A} \textbf{\bibinfo{volume}{52}},
  \bibinfo{pages}{236} (\bibinfo{year}{1996}).

\bibitem[{\citenamefont{Longfield et~al.}(2001)\citenamefont{Longfield,
  Stirling, Lidstr\"om, Mannix, Lander, Stunault, McIntyre, Mattenberger, and
  Vogt}}]{Longfield2001a}
\bibinfo{author}{\bibfnamefont{M.~J.} \bibnamefont{Longfield}},
  \bibinfo{author}{\bibfnamefont{W.~G.} \bibnamefont{Stirling}},
  \bibinfo{author}{\bibfnamefont{E.}~\bibnamefont{Lidstr\"om}},
  \bibinfo{author}{\bibfnamefont{D.}~\bibnamefont{Mannix}},
  \bibinfo{author}{\bibfnamefont{G.~H.} \bibnamefont{Lander}},
  \bibinfo{author}{\bibfnamefont{A.}~\bibnamefont{Stunault}},
  \bibinfo{author}{\bibfnamefont{G.~J.} \bibnamefont{McIntyre}},
  \bibinfo{author}{\bibfnamefont{K.}~\bibnamefont{Mattenberger}},
  \bibnamefont{and} \bibinfo{author}{\bibfnamefont{O.}~\bibnamefont{Vogt}},
  \bibinfo{journal}{Phys. Rev. B} \textbf{\bibinfo{volume}{63}},
  \bibinfo{pages}{134401} (\bibinfo{year}{2001}).

\bibitem[{\citenamefont{Felcher et~al.}(1979)\citenamefont{Felcher, Lander,
  de~V.~du Plessis, and Vogt}}]{Felcher1979}
\bibinfo{author}{\bibfnamefont{G.~P.} \bibnamefont{Felcher}},
  \bibinfo{author}{\bibfnamefont{G.~H.} \bibnamefont{Lander}},
  \bibinfo{author}{\bibfnamefont{P.}~\bibnamefont{de~V.~du Plessis}},
  \bibnamefont{and} \bibinfo{author}{\bibfnamefont{O.}~\bibnamefont{Vogt}},
  \bibinfo{journal}{Solid State Comm.} \textbf{\bibinfo{volume}{32}},
  \bibinfo{pages}{1181} (\bibinfo{year}{1979}).

\bibitem[{\citenamefont{Kuznietz et~al.}(1990)\citenamefont{Kuznietz, Burlet,
  Rossat-Mignod, Vogt, Mattenberger, and Bartholin}}]{Kuznietz1990}
\bibinfo{author}{\bibfnamefont{M.}~\bibnamefont{Kuznietz}},
  \bibinfo{author}{\bibfnamefont{P.}~\bibnamefont{Burlet}},
  \bibinfo{author}{\bibfnamefont{J.}~\bibnamefont{Rossat-Mignod}},
  \bibinfo{author}{\bibfnamefont{O.}~\bibnamefont{Vogt}},
  \bibinfo{author}{\bibfnamefont{K.}~\bibnamefont{Mattenberger}},
  \bibnamefont{and}
  \bibinfo{author}{\bibfnamefont{H.}~\bibnamefont{Bartholin}},
  \bibinfo{journal}{J. Magn. Magn. Mater.} \textbf{\bibinfo{volume}{88}},
  \bibinfo{pages}{109} (\bibinfo{year}{1990}).

\bibitem[{\citenamefont{Vogt and Mattenberger}(1993)}]{Vogt1993}
\bibinfo{author}{\bibfnamefont{O.}~\bibnamefont{Vogt}} \bibnamefont{and}
  \bibinfo{author}{\bibfnamefont{K.}~\bibnamefont{Mattenberger}}, in
  \emph{\bibinfo{booktitle}{Handbook on the Physics and Chemistry of the Rare
  Earths}}, edited by \bibinfo{editor}{\bibfnamefont{K.~A.}
  \bibnamefont{Gschneidner}},
  \bibinfo{editor}{\bibfnamefont{L.}~\bibnamefont{Eyring}},
  \bibinfo{editor}{\bibfnamefont{G.~H.} \bibnamefont{Lander}},
  \bibnamefont{and} \bibinfo{editor}{\bibfnamefont{G.~R.}
  \bibnamefont{Choppin}} (\bibinfo{publisher}{Elsevier (Amsterdam)},
  \bibinfo{year}{1993}), vol.~\bibinfo{volume}{17}, p. \bibinfo{pages}{301}.

\bibitem[{\citenamefont{Blackburn and Bernhoeft}(2006)}]{Blackburn2006a}
\bibinfo{author}{\bibfnamefont{E.}~\bibnamefont{Blackburn}} \bibnamefont{and}
  \bibinfo{author}{\bibfnamefont{N.}~\bibnamefont{Bernhoeft}},
  \bibinfo{journal}{J. Phys. Soc. Jpn.} \textbf{\bibinfo{volume}{75 S}},
  \bibinfo{pages}{63} (\bibinfo{year}{2006}).

\bibitem[{\citenamefont{Lindg{\aa}rd}(2003)}]{Lindgard2003}
\bibinfo{author}{\bibfnamefont{P.~A.} \bibnamefont{Lindg{\aa}rd}},
  \bibinfo{journal}{J. Magn. Magn. Mater.} \textbf{\bibinfo{volume}{266}},
  \bibinfo{pages}{88} (\bibinfo{year}{2003}).

\end{thebibliography}

\end{document}